\documentclass{article}
\usepackage{amsmath,amsfonts,mathrsfs,subfig,graphicx}
\usepackage{nicefrac}
\usepackage{cancel}
\usepackage[T1]{fontenc}
\usepackage[margin=1in]{geometry}

\graphicspath{{./pictures/}{./}}

\newcommand{\re}[1]{(\ref{eq:#1})}

\def\phi{\varphi}
\def\eps{\varepsilon}
\def\rho{\varrho}
\def\d{\mathrm{d}}
\def\p{\partial}

\renewcommand{\vec}[1]{\boldsymbol{#1}}

\newcommand{\rem}[1]{}

\def\={\discretionary{-}{-}{-}}

{\vskip 3pt plus 1pt minus 1pt\begin{normalsize}}%
{\par\end{normalsize}\vskip 3pt plus 1pt minus 1pt}

\newlength{\obrA} 
\newlength{\obrB} 

\newcommand{\Cm}{C\,--\,metric}

\newcommand{\sUP}[3]{\cancel{S}_{(#1,#2)}(#3)}
\newcommand{\sDN}[3]{\bcancel{S}_{(#1,#2)}(#3)}
\newcommand{\sDNs}[3]{\bcancel{S}^2_{(#1,#2)}(#3)}

\begin{document}

\title{Cosmic strings in axisymmetric black hole spacetimes; the \Cm\ ``engines''}

\date{\today}
%\author{David Kofro\v{n}}
%\email{d.kofron@gmail.com}
%\affiliation{
%Institute of Theoretical Physics, Faculty of Mathematics and Physics,\\
%Charles University in Prague,\\
%V Hole\v{s}ovi\v{c}k\'{a}ch 2, 180\,00 Prague 8, Czech Republic}
\author{David Kofro\v{n}\footnote{d.kofron@gmail.com} \\
\emph{\small Institute of Theoretical Physics, Faculty of Mathematics and Physics,}\\
\emph{\small Charles University},\\
\emph{\small V Hole\v{s}ovi\v{c}k\'{a}ch 2, 180\,00 Prague 8, Czech Republic}}

%\pacs{04.20.Jb,04.20.Cv,04.40.Nr,04.70.Bw}
%\keywords{}
\maketitle

\begin{abstract}
The interpretation of so-called cosmic string in black hole spacetimes has settled down to an unsatisfactory state. 

In this article we try to provide a different model for these cosmic strings by explicit construction of these spacetimes from the Bonnor rocket solution. It is shown that the correct stress-energy tensor is that of null dust with a rather strange energy density --- first derivative of Dirac delta distribution.

We will discuss the Schwarzschild solution and the \Cm. In the latter case we will show that there is a momentum flux through the cosmic string, causing the acceleration of the black hole.
\end{abstract}

\section{Introduction}
The cosmic string spacetime --- flat vacuum spacetime with conical singularity and cylindrical symmetry --- can be constructively obtained from a reasonable classic model \cite{Vilenkin_1981}, \cite{Hiscock_1985}. 

In this model the infinite cylinder of matter with density $\rho=\eps$ and negative pressure $p=-\eps$ in one direction only is considered.
Then, the stress energy tensor reads
\begin{align}
T^t_t & = T^z_z = -\eps & \text{for}\ & \rho\leq\rho_0\,, \label{}\\
T^t_t & = T^z_z = 0 & \text{for}\ & \rho>\rho_0\,, \label{}
\end{align}
with all the other components trivially equal to zero.

The Weyl form \cite{Weyl_1917} of static, cylindrically symmetric metric is given by
\begin{equation}
\d s^2 = -e^{2\nu}\,\d t^2 + e^{2\lambda} \left( \d\rho^2+\d z^2 \right)+e^{2\psi} \d\phi^2\,,
\label{eq:}
\end{equation}
where $\nu$, $\lambda$ and $\psi$ are functions of $\rho$ only. 

Then, in the limit when the radius of the cylinder is decreased and the density is increased at the same time so that the ''mass per unit length'' remains constant, a vacuum cosmic string spacetime is constructed, resulting in deficit angle around the axis.

More sophisticated models of cosmic strings have been proposed, see \cite{Christensen_1999,Dyer_1995,Sen_1997}, but all of them consider \emph{cylindrical} symmetry.

Now, the same property --- deficit angle and, thus, the cosmic string --- is inevitably found also in the \Cm\ spacetime and can be implemented in an arbitrary \emph{axisymmetric} solution.

Is it possible to interpret these strings, piercing event horizon and extending to infinity, on the same level as a priori cylindrically symmetric cosmic strings?

To our best knowledge the only attempt to resolve string like sources in GR in general has been done by Israel \cite{Israel77}. His pioneering work suggest to investigate the strings in the near string limit. In this paper Israel himself considered, amongst other examples, static spacetime containing two black holes endowed with cosmic string which keeps them apart in Weyl coordinates. But his near axis limit somehow pushes aside the black hole horizon (which itself is degenerate in Weyl coordinates). 

In order to understand the origin of strings in the Schwarzschild solution or the \Cm\ one should provide a constructive and well controlled procedure. We will do so in the following text. We do not attempt to provide a general treatment of string-like sources.

The Section \ref{sec:math} introduces relevant mathematical definitions employed later in the text and fixes the notation.

In Section \ref{sec:Bonnor} we will briefly review the ``Bonnor rocket'' solution of Einstein field equations which is a quite general black hole solution containing an arbitrary axisymmetric null dust distribution which can vary over the time. The Schwarzschild solution, resp. the \Cm, as a particular example of the Bonnor rocket is treated in Subsection \ref{subsec:Schw}, resp. \ref{subsec:Cm}. Dynamical processes which lead to the string formation or to the smooth transition of Schwarzcshild black hole to the \Cm\ are discussed.

The dynamical situations are difficult to treat, thus, the following Section \ref{sec:st} contains the detailed calculations of the structure of the string in a sequence of static spacetimes.

\section{Mathematical prerequisites} \label{sec:math}
\subsection{Step functions}
We shall use step functions of different profiles. One of them is the non analytic, yet smooth step function $S(x)$
\begin{align}
S(x) &=
\begin{cases}
0 & \text{for\ } x<x_0 \,, \\
\frac{f\left( \frac{x-x_0}{x_1-x_0} \right)}{f\left( \frac{x-x_0}{x_1-x_0} \right)+f\left(  1 - \frac{x-x_0}{x_1-x_0}\right)} &\text{for\ } x\in\langle x_0,\,x_1\rangle \,, \\
1 & \text{for\ } x>x_1 \,,
\end{cases}
\label{}
\end{align}
where $f(x) = e^{-\nicefrac{1}{x}}$, which is a step in between $x_0$ and $x_1$.

The another class of step functions, so called smoothstep functions, are polynomials of order $2n+1$ with boundary conditions $f(x_0)=0$, $f(x_1)=1$ and $\d^j f(x_0) / \d x^j = \d^j f(x_1) / \d x^j = 0$  for $j=1,\,2,\dots n$ which are simple to construct, manage analytically and are smooth up to the order $n$.

In general we will have a step up function $\sUP{a}{b}{x}$ which vanishes for $x<a$, has a desired interpolation in between 0 and 1 for $x\in\langle a,\,b\rangle$ and is equal 1 for $x>b$. The step down is then simply $ \sDN{a}{b}{x} = 1-\sUP{a}{b}{x}$. 

Also the ``table'' function 
\begin{align}
T_{(a,b,c,d)}(x)\quad &=\quad \begin{cases}
\quad  0                &\text{for } x<a \,,\\
\quad  \sUP{a}{b}{x}  &\text{for } x\in\langle a,\,b\rangle \,,\\
\quad  1                &\text{for } x\in\langle b,\,c\rangle \,,\\
\quad  \sDN{c}{d}{x}  &\text{for } x\in\langle c,\,d\rangle \,,\\ 
\quad  0                &\text{for } x>d \,,\\
 \end{cases} 
\label{eq:Table}
\end{align}
will be of use.

\subsection{Fourier\,--\,Legendre series} \label{subsec:Legendre}
$L^2$ functions on the interval $\langle-1,1\rangle$ can be expanded in the basis of Legendre polynomials as
\begin{equation}
f(x) = \sum_{n=0}^{\infty} a_n P_n(x) \,,
\label{eq:}\end{equation}
where, due to the normalization of Legendre polynomials, the coefficients $a_n$ are
\begin{equation}
a_n = \frac{2n+1}{2}\, \int_{-1}^{1} f(x) P_n(x)\, \d x \,.
\label{eq:}
\end{equation}
The expansion of the Dirac delta distribution is given by (\cite{NIST} (1.17.22))
\begin{equation}
\delta (x-a) = \sum_{n=0}^{\infty} \left( n+\nicefrac{1}{2} \right) P_n(x) \, P_n(a) \,.
\label{eq:deltaFL}
\end{equation}

Legendre polynomials arise as the result of Gramm\,--\,Schmidt orthogonalization of monomials $\{x^j,\,j=0\ldots \infty\}$. In our calculations we needed the inverse relation
\begin{equation}
x^k = \sum_{j=0}^k a^{(k)}_j P_j(x) \,,
\label{eq:}\end{equation}
whose explicit formulae (which differ for even and odd powers of $x$) we found to be
\begin{align}
x^{2k} & = \sqrt{\pi}\; \frac{\Gamma(2k+1)}{2^{2k+1}} \;
           \sum_{j=0}^{k} \frac{4j+1}{\Gamma(k+j+\nicefrac{3}{2})\, \Gamma(k-j+1)}\;P_{2j}(x)\,, \\
x^{2k+1} &= \sqrt{\pi}\; \frac{\Gamma(2k+2)}{2^{2k+2}} \;
            \sum_{j=0}^{k} \frac{4j+3}{\Gamma(k+j+\nicefrac{5}{2})\, \Gamma(k-j+1)}\;P_{2j+1}(x)\,.
\label{eq:invLeg}
\end{align}

\section{Bonnor rocket} \label{sec:Bonnor}
In 1996 Bonnor \cite{Bonnor_1996} found a solution of Einstein field equations with null dust in which the central black hole can radiate the null dust with an arbitrary axisymmetric pattern and time profile. The modern version of this metric can be found in \cite{GrPoB} and reads as follows
\begin{multline}
\d s^2 = -2\,\d u\,\d r - \left( -\frac{1}{2}\,G_{,xx}-\frac{2m(u)}{r}-r\left( bG \right)_{,x}-b^2Gr^2 \right)\d u^2 %\\
 + 2br^2\,\d u\,\d x + r^2\left( \frac{\d x^2}{G}+G\,\d\phi^2 \right),
\label{eq:BR}
\end{multline}
where
\begin{align}
b(x,u) &= -A(u) -\int\frac{G_{,u}(x,u)}{G^2(x,u)}\,\d x\,, \label{eq:b}\\
G(x,u) &= \left( 1-x^2 \right)\left[ 1+\left( 1-x^2 \right)h(x,u) \right], 
\label{eq:origG}
\end{align}
with $A(u)$ an arbitrary function of $u$ and $h(x,u)>-1$ an arbitrary smooth bounded function.
This represents a Bonnor rocket, a particle emitting null dust (pure radiation) with following angular dependence
\begin{equation}
4\pi\, n^2(x,u) = -\frac{1}{8}\,\left( GG_{,xxx} \right)_{,x}+\frac{3}{2}\,m\left( bG \right)_{,x}-m_{,u} \,.
\label{eq:n}
\end{equation}
Corresponding stress energy tensor reads
\begin{align}
T_{ab} &= \rho\, l_a l_b  \,, & 
\rho   &= \frac{n^2}{r^2} \,, &
l^a    &= \frac{\p}{\p r} \,.
\label{eq:}
\end{align}

The form of $G(x,\,u)$ given by \re{origG} is not the most general one. It has been chosen by Bonnor so that the axis is regular. And, clearly, it does not guarantee that the quantity $n^2$ defined in \re{n} is positive (while the null energy conditions seems to be a minimal physical requirement).

Let us relax these restrictions, first of all we will consider the function 
\begin{equation}
G(x,\,u) = \left( 1-x^2 \right)\left( 1+ h(x,\,u) \right)
\label{eq:}
\end{equation}
and then we will omit the second power in the definition of $n(x)$.

Investigating regularity condition \cite{Stephani2003} of the axis given in terms of the norm of the axial Killing vector $\vec{\xi}_{(\phi)}$
\begin{equation}
\frac{1}{4} \lim_{x\rightarrow \pm 1} \frac{F_{,a}F^{,a}}{F} = 1+h(\pm 1,\,u)\,,\qquad \text{where} \qquad F=\vec{\xi_{(\phi)}}\cdot\vec{\xi_{(\phi)}}
\label{eq:}
\end{equation}
it is clear the function $h(x,\,u)$ determines the regularity of the axis.

The Bonnor rocket \re{BR}\,--\,\re{origG} was fine tuned and thus in its original form does not contain conical singularities. We can introduce them by rescaling $G\rightarrow KG$ which is equivalent to the choice $h(x,\,u)=\,$const. 
 
Then, scaling the coordinates as 
\begin{align}
\tilde{u}& = \sqrt{K}\,u \,, &
\tilde{r}& = r/\sqrt{K} \,, &
\tilde{m}& = m/K\sqrt{K} \,, &
\tilde{A}& = A/\sqrt{K}\,,
\end{align}
leads to the same form \re{BR} the metric, except the term $K^2\d\phi^2$ which shows the presence of conical singularity as we consider $\phi$ to run form $0$ to $2\pi$ strictly.

This ``relaxed'' class of Bonnor rockets contains Schwarzschild solution even with conical singularities as well as the \Cm.

\subsection{The Schwarzschild solution} \label{subsec:Schw}
The metric \re{BR} with $b=0$ and $G=(1-x^2)$ is the Schwarzschild solution. For $b=0$ and $G=K(1-x^2)$ and after the aforementioned rescaling the Schwarzschild black hole with the horizon pierced by cosmic string is obtained. 

\begin{figure}
\begin{center}
\subfloat[Schwarzschild]{\includegraphics[width=\obrB,keepaspectratio]{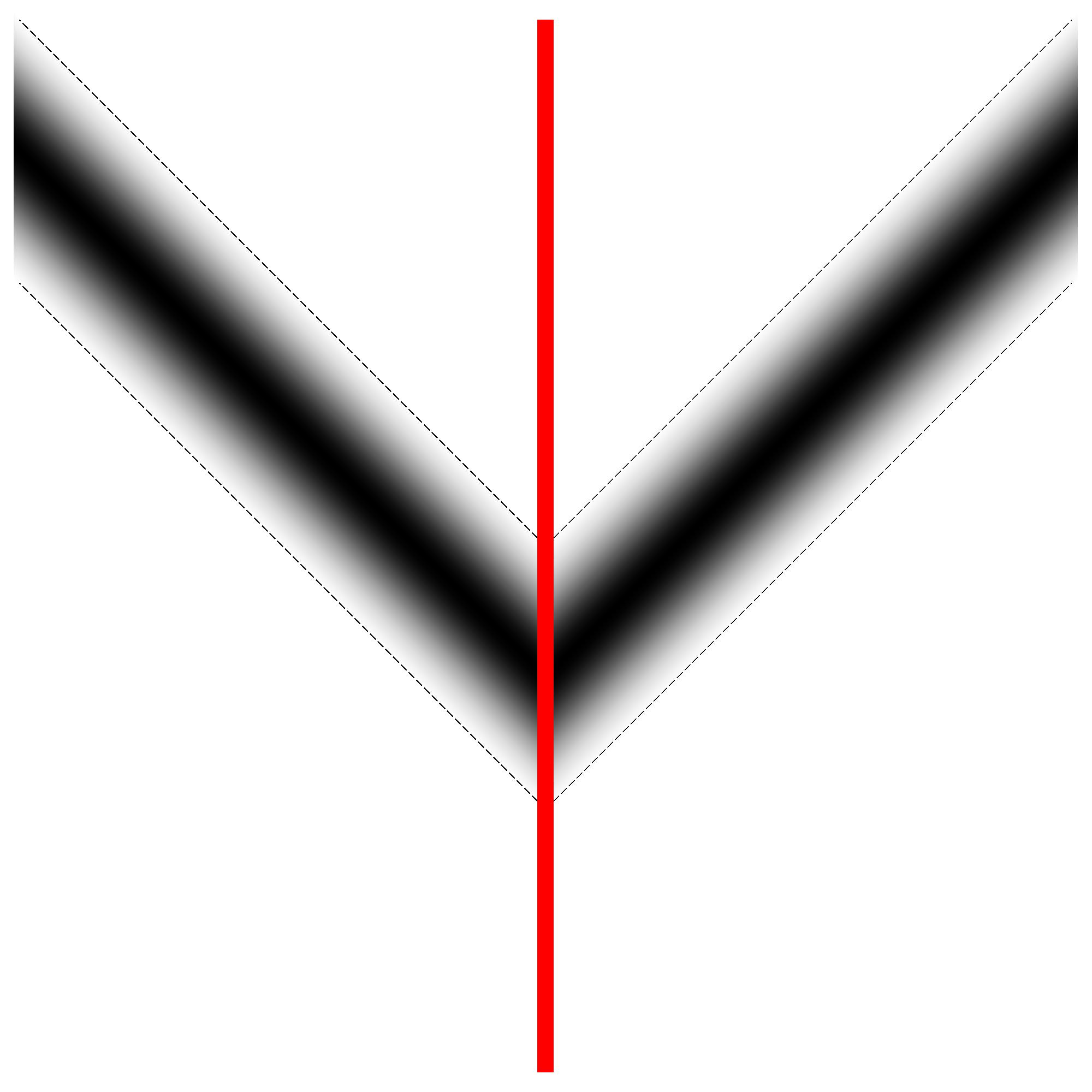}} \qquad
\subfloat[Schwarzschild $\longrightarrow$ \Cm]{\includegraphics[width=\obrB,keepaspectratio]{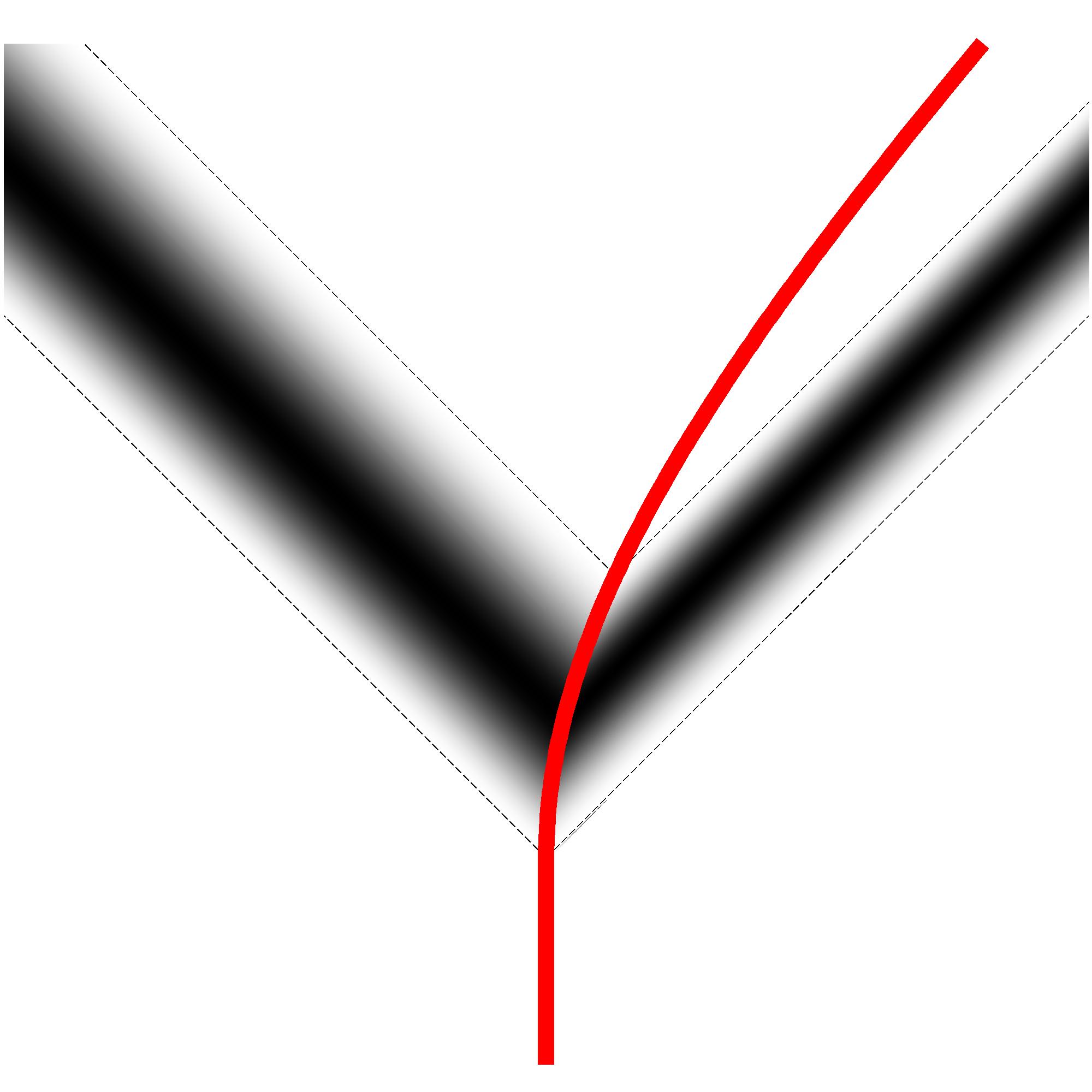}}
\end{center}
\caption{Dynamical formation of the cosmic string in the Schwarzschild solution (a) and the smooth transition (nonuniform acceleration) of  the Schwarzschild black hole to the \Cm, accompanied with the formation of cosmic strings with different stress energy tensors along the north and south poles.\\ 
The black hole horizon is depicted by bold red line. At $u=u_0$ the radiation phase starts, the gray scale represents the time evolution of the intensity of the radiation, and at $u=u_1$ the spacetime becomes static again, with conical singularities present.}
\label{fig:acc}
\end{figure}

Using
\begin{equation}
G(x,\,u) = \left( 1-x^2 \right)\left( 1+ 2w \sUP{u_0}{u_1}{u} e^{-\frac{\sDN{u_0}{u_1}{u}}{1-x^2}} \right)
\label{eq:}
\end{equation}
we get a transition between Schwarzschild for ($u<u_0$) through a radiating phase $u\in\langle u_0,\, u_1\rangle$ during which the axis is still regular, to a Schwarzschild pierced by cosmic string with $K=1+2w$ which appears at $u=u_1$ and there is no evolution later, see Figure \ref{fig:acc} (a) for a  schematic picture.

\begin{figure}
\centering
\subfloat[]{\includegraphics[width=\obrB,keepaspectratio]{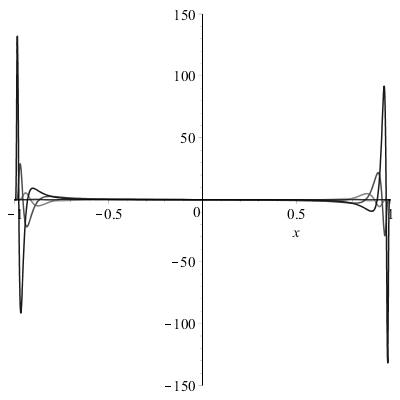}} \qquad
\subfloat[]{\includegraphics[width=\obrB,keepaspectratio]{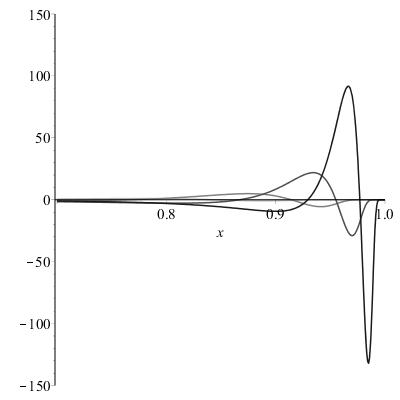}} 
\caption{An example of radiation pattern and its focusing properties. The profile is calculated for the structure function \re{expCm}, as the limiting parameter $\eps$ approaches zero, the beam of null radiation gets focused along the axis (the smaller $\eps$ the darker the line).}
\label{fig:rp}
\end{figure}

Investigating the radiation pattern \re{n} we can see that the radiation gets more and more focused (see Figure \ref{fig:rp} for an example) until the string appears and propagates to the infinity along a null worldline. 

\subsection{The \Cm} \label{subsec:Cm}
The \Cm\ in Robinson\,--\,Trautman coordinates \cite{GrPoB} (Eq. 19.4 therein) reads 
\begin{equation}
\d s^2 = -2\,\d u\,\d r +A^2r^2G\left( x-\frac{1}{Ar} \right) \d u^2 - 2Ar^2\, \d u\, \d x + r^2\left( \frac{\d x^2}{G(x)}+G(x)\,\d\phi^2 \right),
\label{eq:cm}
\end{equation}
with
\begin{equation}
G(x) = \left(1-x^2\right)\left( 1+2Amx \right).
\label{eq:}
\end{equation}

Clearly, the metric element \re{BR} of the Bonnor rocket contains the \Cm\ \re{cm} as a special case, we simply have to set
\begin{align}
m(u) &=m\,, & 
b(x,u) &= -A\,, &
G(x,u) &= \left( 1-x^2 \right)\left( 1+2Amx \right).
\label{}
\end{align}
Choosing the functions $G(x,\,u)$ and $A(u)$ in general Bonnor rocket metric \re{BR} as
\begin{align}
G(x,\,u) &= \left( 1-x^2 \right)\left( 1+2Am x \,\sUP{u_0}{u_1}{u}e^{-\frac{\sDN{u_0}{u_1}{u}}{1-x^2}} \right) \,, \\
A(u) &= A\sUP{u_0}{u_1}{u} \,,
\label{eq:Gg}
\end{align}
we get a smooth transition from the static Schwarzschild solution for $u<u_0$, through a dynamic radiation phase for $u\in\langle u_0,\,u_1\rangle$ during which the radiation gets more a more focused along the still regular axis (but this time this radiation pattern is not reflectively symmetric) to the \Cm\ for $u>u_1$. See Figure \ref{fig:acc} (b) for schematic picture. The axis start to posses a conical singularity at $u=u_1$ when the radiation is completely focused into an infinitely narrow beam and this singularity propagates along the null direction to infinity.

This dynamically obtained \Cm\ is for $u>u_1$ diffeomorphic to the \Cm\ but, clearly, cannot be analytically extended and does not contain the second black hole accelerated in opposite direction.

The term $\int G(x,u)_{,u} / G^2(x,u)\, \d x$ hidden in the definition of the function $b(x,\,u)$ and thus in the radiation pattern $n(x,\,u)$, see \re{n}, is difficult, even impossible, to threat analytically. Therefore, in the next section, we will investigate these spacetimes as a sequence of different static spacetimes parameterized either by continuous parameter $\epsilon$ or integer $N$.

\begin{figure}
\centering
\includegraphics[width=\obrA,keepaspectratio]{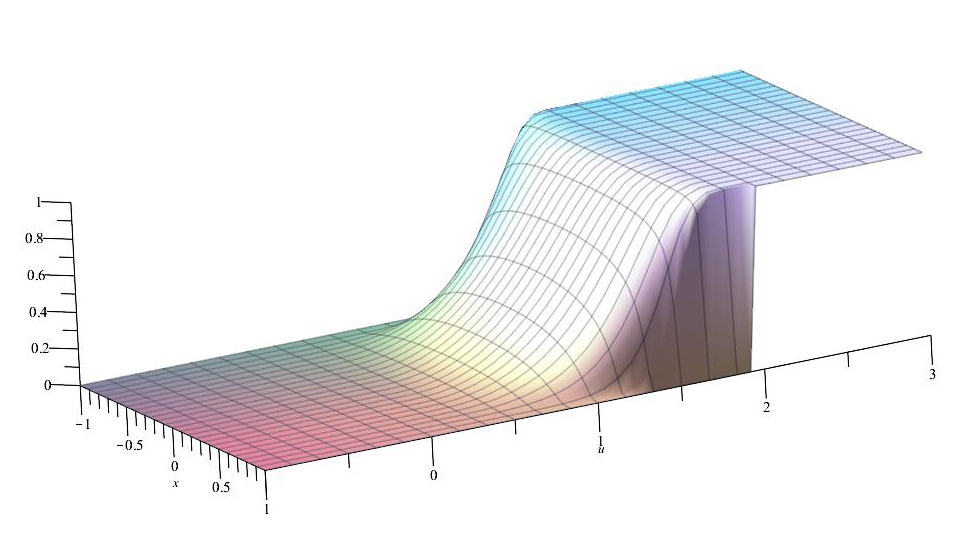}
\caption{The function $\sUP{u_0}{u_1}{u}\, e^{-\frac{\sDN{u_0}{u_1}{u}}{1-x^2}}$ which enters the structure function $G(x,\,u)$ and changes it dynamically in between $u_0$ and $u_1$.}
\label{fig:step}
\end{figure}

\section{Static treatment} \label{sec:st}
The Bonnor rocket emits null radiation and thus loses its mass, given by the energy outflow
\begin{equation}
\oint r^2 \rho\, \d\Omega = \oint n(x)\, \d\Omega \,.
\label{}
\end{equation}
Even if there is no time evolution there can be pure radiation, therefore our demand for staticity requires
\begin{align}
\oint n(x)\, \d \Omega &= 0\,, & m(u) &=m \,.
\label{eq:stac}
\end{align}
From this follows that $n(x)$ cannot be positive for $\forall x \in \langle -1,\,1\rangle$.

The arbitrariness in the choice of function $h(x)$ is almost infinite. Let us consider a sequence of spacetimes labelled by $N\in(\mathbb{N}\cup 0)$ given by
\begin{align}
G(x,u) &= \left( 1-x^2 \right)\left( 1+2w\left( 1-x^{2N} \right) \right),
\label{eq:Schw-N}
\end{align}
which can be treated completely analytically.

For $N=0$ we get a standard Schwarzschild solution while in the limit $N\rightarrow \infty$ the Schwarzschild solution with a cosmic string is obtained. During the limiting process the axis is regular all the time.
 
Evaluating the radiation pattern \re{n} for the structure function \re{Schw-N} leads trivially to  
\begin{align}
4\pi\, n_N(x) = &       -2 w^2\, (4N+1)(2N+1)(N+1)N\, x^{4N} %\nonumber \\
                       +4 w^2\, (2N^2+1)(4N-1)N\, x^{4N-2} \nonumber  \\
                &       -2 w^2\, (4N-3)(2N-1)N(N-1)\, x^{4N-4} %\nonumber \\
                     +(2w+1)w\, (2N+1)^2\,(N+1)N\, x^{2N}  \\
                &    -2(2w+1)w\, (2N^2+1)(2N-1)N\, x^{2N-2} %\nonumber \\
                     +(2w+1)w\, (2N-1)(2N-3)N(N-1)\, x^{2N-4}\nonumber \,.
\label{eq:rpN}
\end{align}

In this explicit and exact form the monomials $x^j$ can be expressed (or expanded) in the basis of Legendre polynomials as shown in the Section \ref{subsec:Legendre}. Then the limit $N\rightarrow \infty$ of $n_N(x)$ leads to  
\begin{equation}
4\pi\, n(x) = \left( w^2+w \right) \sum_{n=0}^{\infty} 2n\left( 2n+\nicefrac{1}{2} \right)\left( 2n+1 \right) P_{2n}(x) \,.
\label{eq:serSchw}
\end{equation}
This series can be summed up using the expansion of Dirac delta distribution \re{deltaFL} in the basis of Legendre polynomials from which we get
\begin{equation}
\Delta_+ \equiv \delta(x+1) + \delta(x-1) = 2\sum_{n=0}^{\infty} \left( 2n+\nicefrac{1}{2} \right)P_{2n}(x) \,.
\label{eq:}
\end{equation}
Now, employing the standard properties of Legendre polynomials we get
\begin{align}
\frac{1}{2}\, \frac{\d}{\d x}\left[ \left( 1-x^2 \right)\frac{\d}{\d x}\,\Delta_{+} \right] & =-\frac{\d}{\d x}\,\delta(x+1) +\frac{\d}{\d x}\,\delta(x-1) \nonumber \\
& =  -\sum_{n=0}^{\infty} 2n\left( 2n+\nicefrac{1}{2} \right)\left( 2n+1 \right) P_{2n}(x)\,,
\label{eq:}
\end{align}
and thus the final radiation pattern is
\begin{equation}
4\pi\,n(x) = -(w^2+w)\left[ \frac{\d}{\d x}\,\delta(x+1) -\frac{\d}{\d x}\,\delta(x-1) \right] .
\label{eq:DiracSchw}\end{equation}
This leads us to one of the main results of this paper -- to the explicit form of stress energy tensor for the cosmic string piercing the Schwarzschild black hole
\begin{equation}
T_{ab} = -(w^2+w)\left[ \frac{\d}{\d x}\,\delta(x+1) -\frac{\d}{\d x}\,\delta(x-1) \right] \frac{l_al_b}{r^2}\,.
\end{equation}

Analogously, the \Cm\ can be obtained as a limiting case of the following sequence of spacetimes
\begin{align}
G(x,u) & = \left( 1-x^2 \right)\left( 1+2Amx\left( 1-x^{2N} \right) \right), \\
A(u) &= A\left( 1-\frac{1}{N+1} \right),
\label{}
\end{align}
for which the condition \re{stac} of zero mass flux through an arbitrary sphere holds. Evaluating the radiation pattern is straightforward (but not short enough to be presented). Expressing monomials in the basis of Legendre polynomials and taking the limit $N\rightarrow \infty$ yields
\begin{align}
4\pi\, n(x) = & A^2m^2 \sum_{n=0}^\infty 2n \left( 2n+\nicefrac{1}{2} \right)\left( 2n+1 \right) P_{2n}(x) %\nonumber \\
 + Am \sum_{n=0}^\infty \left( 2n+1 \right)\left( 2n+\nicefrac{3}{2} \right)\left( 2n+2 \right) P_{2n+1}(x) \,.
\label{eq:serCm}
\end{align}
After some rearrangement of the expansion of the Dirac delta distribution in Legendre polynomials,
\begin{equation}
\Delta_- \equiv \delta(x+1) - \delta(x-1) = 2\sum_{n=0}^{\infty} \left( 2n+\nicefrac{3}{2} \right)P_{2n+1}(x) \,,
\label{eq:}
\end{equation}
and employing the properties of Legendre polynomials again we get
\begin{align}
\frac{1}{2}\, \frac{\d}{\d x}\left[ \left( 1-x^2 \right)\frac{\d}{\d x}\,\Delta_{-} \right] & =-\frac{\d}{\d x}\,\delta(x+1) -\frac{\d}{\d x}\,\delta(x-1) \nonumber\\
& =  -\sum_{n=0}^{\infty} \left( 2n+1 \right)\left( 2n+\nicefrac{3}{2} \right)\left( 2n+2 \right) P_{2n+1}(x)\,.
\label{eq:}
\end{align}
As a result, we can recognize \re{serCm} to be 
\begin{align}
%n(x)= &\,-A^2m^2 \left( \frac{\d}{\d x}\, \delta (x+1) - \frac{\d}{\d x}\, \delta(x-1) \right)\nonumber \\
%& - Am \left( \frac{\d}{\d x}\, \delta(x+1) + \frac{\d}{\d x}\, \delta(x-1) \right) \\
4\pi\, n(x) = & -Am\left( Am+1 \right)\frac{\d}{\d x}\,\delta(x+1) 
  + Am\left( Am-1 \right)\frac{\d}{\d x}\,\delta(x-1) \,.
\label{eq:DiracCM}\end{align}

In the case of the \Cm\ the null dust is not radiated away in a symmetric manner and thus carries the momentum away, in the rest frame of the black hole we find
\begin{equation}
P_z = \int_{-1}^1 n(x) x\, \d x = Am \,,
\label{eq:Pz}
\end{equation}
where $x$ is spherical harmonics $Y^0_1(x,\phi)$ and actually should be replaced by the solution of eigenfunctions on two sphere $t=\,$const and $r=\,$const as we have done in \cite{Kofron_2015, Kofron_2016}. Unfortunately the solution can be found only in terms of Heun general function and cannot be normalized. Yet, this solution for small $Am$ tend to $x$ and the corrections to $P_z$ given by \re{Pz} would be of order $A^2m^2$.

A different profile whose advantages lie in the fact that for $x\in\langle-1+\eps,\,1-\eps\rangle$ the spacetime is locally Schwarzchild or the \Cm\ can be found
\begin{align}
G(x,u) &= \left( 1-x^2 \right)\left( 1+2w \sUP{-1}{0}{-\eps} T_{(-1,-1+\eps,1-\eps,1)} (x) \right) \,,
\label{eq:Schw-e}
\end{align}
for the Schwarzchild or
\begin{align}
G(x,u) &= \left( 1-x^2 \right)\left( 1+2Am x \sUP{-1}{0}{-\eps} T_{(-1,-1+\eps,1-\eps,1)} (x) \right) \,,
\label{eq:Cm-e}
\end{align}
for the \Cm. Step functions are now the polynomial smoothstep of order 7 or higher. In these cases we can calculate $n_\eps(x)$ and then, using computer algebra systems, its Fourier\,--\,Legendre expansion. In the next step --- in the limit $\eps \rightarrow 0^+$ we recover the results \re{serSchw} and \re{serCm}.

A completely different approach, which shows that these results are robust, is to use the functions
\begin{align}
G_\eps (x,u) &= \left( 1-x^2 \right)\left( 1+ 2w  e^{-\frac{\eps}{1-x^2}} \,\sDN{0}{1}{\eps} \right)\,, & A(u) & = 0\,, 
\end{align}
for Schwarzschild
\begin{align}
G_\eps (x,u) &= \left( 1-x^2 \right)\left( 1+2Am  e^{-\frac{\eps}{1-x^2}}\,\sDN{0}{1}{\eps} \right)\,, & A(u) &= A\sDN{0}{1}{\eps}\,,
\label{eq:expCm}
\end{align}
for \Cm.

Evaluating the radiation pattern $n_\eps (x)$ is straightforward but it is impossible to express this function in the Fourier\,--\,Legendre series due to the integrals --- they consist of rational function multiplied by $e^{-\nicefrac{\eps}{(1-x^2)}}$.

For the Schwarzschild solution we get
\begin{align}
4\pi\, n_\eps (x) &= 
   -2w^2\sDNs{0}{1}{\eps}\, \frac{p_8(x)}{(x-1)^6(x+1)^6}\,\eps^2\,e^{-\frac{2\eps}{1-x^2}} %\nonumber \\ 
    -w\sDN{0}{1}{\eps}\, \frac{q_8(x)}{(x-1)^6(x+1)^6}\,\eps^2\,e^{-\frac{\eps}{1-x^2}}\,, 
\label{}
\end{align}
where $p_8(x)$ and $q_8(x)$ are polynomials
\begin{align}
p_8(x) &= 8\eps^2 x^4 - 2x^2\left( 1-x^2 \right)\left( 11x^2+9 \right)\eps +3\left( 3x^4+8x^2+1 \right)\left( 1-x^2 \right)^2 , \\
q_8(x) &= 4\eps^2 x^4 - 2x^2\left( 1-x^2 \right)\left( 4x^2+3 \right)\eps +3\left( 3x^4+8x^2+1 \right)\left( 1-x^2 \right)^2 .
\label{}
\end{align}
Therefore, for now, consider the $n_\eps (x)$ as a distribution and let it act on test functions. We anticipate the result, of course. The behavior of $n_\eps (x)$ is governed by the term $e^{-\nicefrac{\eps}{(1-x^2)}}$, for $x\in (-1,\,1)$ the limit $\eps\rightarrow 0^+$ tends to 0.

In the radiation pattern $n_\eps (x)$ we can interpolate for small $\eps$ 
\begin{align}
e^{-\frac{\eps}{1-x^2}} & \sim e^{-\frac{1}{2}\frac{\eps}{x+1}}\,, &\text{for}\ & x\in\langle -1, 1);\ \text{denoted by}\ n^-_\eps\\
e^{-\frac{\eps}{1-x^2}} & \sim e^{\frac{1}{2}\frac{\eps}{x-1}}\,, &\text{for}\ & x\in (-1,\, 1\rangle;\ \text{denoted by}\ n^+_\eps 
\label{}
\end{align}
and similarly for $e^{-\nicefrac{2\eps}{(1-x^2)}}$. 

Using computer algebra systems it can be analytically calculated how this distribution acts on basis of polynomials, i.e. evaluate the integral $n_\eps(x)\left[ x^N \right] = \int_{x_0}^1 x^N n_\eps(x)$. In the limit the result, independent on $x_0 \in (-1,\,1)$, is
\begin{align}
\lim_{\eps\rightarrow 0^+} \int_{x_0}^1 x^N n^+_\eps(x)\, \d x & = (w^2+w)N = -(w^2+w)\, \delta'(x+1) \left[ x^N \right]\,,\\
\lim_{\eps\rightarrow 0^+} \int_{-1}^{x_0} x^N n^-_\eps(x)\, \d x & = (-1)^N N(w^2+w) = -(w^2+w)\, \delta'(x-1)\left[ x^N \right]\,,
\label{eq:}
\end{align}
and thus it acts as derivative of Dirac delta distribution as in \re{DiracSchw}.

The same procedure can be repeated for the \Cm\ with results as in \re{DiracCM}, of course.

\section{Israel's approach}
The natural question which arises is why these results are different to those well accepted and agreed on in the literature which are founded on the Israel paper \cite{Israel77}.

Israel's approach is based on explicit construction of coordinates in the vicinity of the axis such that the metric is of the form
\begin{equation}
\d s^2 = \d \rho^2 + A^2(z,t)\d z^2 + B^2 \rho^2\d\phi^2 - C^2(z,t)\d t^2 \,.
\label{eq:}
\end{equation}
The extrinsic curvature $K_{ab}$ (and its densitised form $\mathscr{K}_a^b$) of cylinders of constant $\rho$
\begin{align}
K_{ab} & = \frac{1}{2}\,\frac{ \p g_{ab} }{\p\rho}\,, &
\mathscr{K}_a^b & = \sqrt{-\det g}\, K_a^b\,,
\label{eq:exc}
\end{align}
and its limit
\begin{equation}
\mathscr{C}_a^b = \lim_{\rho\rightarrow 0^+} \mathscr{K}_a^b \,.
\label{eq:}\end{equation}

Let us have a Schwarzchild solution endowed with cosmic string in Weyl coordinates
\begin{equation}
\d s^2 = -e^{2\psi}\, \d t^2 + e^{2\left(\lambda-\psi\right)}\, \left( \d r^2 + \d z^2 \right)+e^{-2\psi}\,r^2\,\d\phi^2 \,,
\label{eq:}\end{equation}
where
\begin{align}
\psi &= \frac{1}{2}\ln \left[ \frac{R_++R_--2m}{R_++R_-+2m} \right], &
\lambda &= \frac{1}{2}\ln \left[ \frac{\left( R_++R_- \right)^2-4m^2}{4R_+R_-} \right] + K \,,
\label{}
\end{align}
with $R_\pm = \sqrt{r^2+\left( z\pm m \right)^2}$. The parameter $K$ controls the regularity of the axis --- regularity condition reads $\lambda(0,z)=0$.

We can find the transformation from Weyl coordinates $(r,\,z)$ to approximate coordinates $(\rho,\,\zeta)$ in which $\rho$ is the affine parameter of geodesic connecting the axis with the point in its vicinity to an arbitrary order of precision (for $\zeta>m$) 
\begin{align}
r &= 0 
   + e^{-K}\sqrt{\frac{\zeta-m}{\zeta+m}}\, \rho 
   + \frac{e^{-3K}}{6}\frac{\sqrt{\zeta^2-m^2}m}{\left( \zeta+m \right)^4}\, \rho^3 %  \nonumber \\ 
   - \frac{e^{-5K}}{120} \frac{\sqrt{\left( \zeta^2-m^2 \right)}m(2m+9\zeta)}{\left( \zeta+m \right)^7}\, \rho^5 + \dots \,, \\
z &= \zeta 
   - \frac{e^{-2K}}{2}\frac{m}{\left( \zeta+m \right)^2}\, \rho^2 
   - \frac{e^{-4K}}{24}\frac{m\left( m-3\zeta \right)}{\left( \zeta+m \right)^5} \rho^4% \nonumber \\
    + \frac{e^{-6K}}{720}\frac{m\left( 35m^2+6m\zeta-45\zeta^2 \right)}{\left( \zeta+m \right)^8}\, \rho^6 + \dots \,,
\label{}
\end{align}
and then the Schwarzschild metrics reads
\begin{align}
\d s^2 & = -\frac{\zeta-m}{\zeta+m}\left( 1 + me^{-2K}\,\frac{1}{\left(\zeta+m\right)^3}\,\rho^2-\frac{me^{-4K}}{12}\frac{ 13m-9\zeta }{\left( \zeta+m \right)^6}\,\rho^4 + o(\rho^6) \right) \d t^2 \nonumber \\
 & +\left( 1+ o(\rho^6)\right) \d \rho^2  + o(\rho^7)\, \d\rho\,\d\zeta \nonumber\\
 & +\frac{\zeta+m}{\zeta-m}\left( e^{2K} + \frac{m}{\left( \zeta+m \right)^3}\,\rho^2+\frac{me^{-2K}}{12}\frac{7m-3\zeta}{\left( \zeta+m \right)^6}\,\rho^4 + o(\rho^6) \right) \d\zeta^2 \nonumber \\
 & + e^{-2K}\rho^2 \left( 1-\frac{2}{3}\frac{e^{-2K}}{\left( \zeta+m \right)^3}\,\rho^2 + o(\rho^4) \right) \d\phi^2 \,.
\label{}
\end{align}

The limit of densitised external curvature tensor is simply 
\begin{align}
\mathscr{C}_a^b&= \begin{pmatrix} 0 & 0 & 0 \\ 0 & 0 & 0 \\ 0 & 0 & 1 \end{pmatrix} \,,
\end{align}
in coordinates $(t,\,\zeta,\,\phi)$. But this tensor has null eigenvectors; and therefore the Israel's approach does not need to work, as \emph{``Condition (vi) excludes ``lightlike'' sources which (like null surface layers) require special treatment''}. Although the axis itself is not a null hypersurface, the stress energy tensor is composed of null dust.

We have already seen in Section \ref{subsec:Schw} and \ref{subsec:Cm} that the nature of singularities is lightlike.

This shows that the conical defects are a very subtle subject which has to be treated carefully.

\section{Conclusions}
We have shown that the conical singularity in the black hole spacetimes represents a cosmic string with a different stress energy tensor than is usually accepted.

Also, by explicit construction, we revealed the structure of the conical singularities piercing black hole horizons and calculated momentum flux through these.

{\bf Acknowledgements} D.K. acknowledges the support from the Czech Science Foundation, Grant 17-16260Y. Moreover, D.K. would like to thank Dr. M. Scholtz for inspiring discussions and comments on the manuscript.

\bibliography{kofron}{}
\bibliographystyle{plain}

\end{document}